# Cost-Driven Offloading for DNN-based Applications over Cloud, Edge and End Devices

Bing Lin, Yinhao Huang, Jianshan Zhang, Junqin Hu, Xing Chen, Jun Li

*Abstract*—Currently, deep neural networks (DNNs) have achieved a great success in various applications. Traditional deployment for DNNs in the cloud may incur a prohibitively serious delay in transferring input data from the end devices to the cloud. To address this problem, the hybrid computing environments, consisting of the cloud, edge and end devices, are adopted to offload DNN layers by combining the larger layers (more amount of data) in the cloud and the smaller layers (less amount of data) at the edge and end devices. A key issue in hybrid computing environments is how to minimize the system cost while accomplishing the offloaded layers with their deadline constraints. In this paper, a self-adaptive discrete particle swarm optimization (PSO) algorithm using the genetic algorithm (GA) operators was proposed to reduce the system cost caused by data transmission and layer execution. This approach considers the characteristics of DNNs partitioning and layers offloading over the cloud, edge and end devices. The mutation operator and crossover operator of GA were adopted to avert the premature convergence of PSO, which distinctly reduces the system cost through enhanced population diversity of PSO. The proposed offloading strategy is compared with benchmark solutions, and the results show that our strategy can effectively reduce the system cost of offloading for DNN-based applications over the cloud, edge and end devices relative to the benchmarks.

## I. INTRODUCTION

CONTEMPORARILY, deep neural networks (DNNs) have achieved a great success in various applications, such as natural language processing [1], speech recognition [2], and computer vision [3]. For example, Fig. 1 shows the state-of-the-art performances continuously achieved by DNNs in computer vision. Meanwhile, the number of Internet of Things (IoT) devices has increased dramatically. These end devices, equipped with sensors (*e.g.*, microphones, cameras, and gyroscopes) for obtaining a large amount of environment data, are usually attractive to machine learning (ML) applications [4].

However, the IoT devices with limited energy and computing resources cannot afford computation-intensive tasks (*e.g.*, DNNs). Performing classification directly on the IoT devices by simple ML model leads to low system accuracy [5]. As such, DNNs are conventionally deployed in the cloud with powerful computation capability. This results in a prohibitively serious delay when offloading input sensor data to DNNs in the cloud, due to the long distance between the cloud and IoT devices.

Mobile edge computing (MEC) is proposed as a promising computing model for solving the problem by deploying servers at the network edge close to the end devices [6]. One solution to reducing the system delay of offloading for DNN-based applications is to partition DNNs [7] in hybrid computing environments, consisting of the cloud, edge, and end devices, and combine the larger layers (more amount of data) in the cloud and the smaller layers (less amount of data) at the edge and end devices. In this way, the traffic load of core network and the transmission delay will be alleviated significantly, and the overall system accuracy will be improved [5].

Offloading for DNN-based applications in MEC has been broadly studied [7-11]. These work mostly focuses on offloading DNN layers to the edge instead of to the cloud. However, much less attentions are paid to offloading DNN layers in hybrid computing environments [5]. It is a challenging task to partition DNNs and schedule different layers to their suitable locations for distinct applications while satisfying each application's deadline constraint. Moreover, when the input/intermediate data is scheduled to different cloud/edge servers, the cost of computation and transmission are also different. Therefore, how to minimize the system cost while accomplishing the offloaded layers within their deadlines in hybrid computing environments is still an open issue.

In our previous work [12, 13], we addressed the time-driven data placement for a scientific workflow combining edge computing and cloud computing, as well as the cost-driven scheduling for deadline-based workflow across multiple clouds, respectively. Offloading DNN layers in hybrid computing environments and workflow scheduling across multiple clouds are both NP-hard problems with many similarities, such as the structure between the DNNs and the workflow [14]. In this paper, we propose a self-adaptive particle swarm optimization (PSO) algorithm using the genetic algorithm (GA) operators (PSO-GA) to reduce the system cost caused by data transmission and layer execution, with the deadline constraints of all DNN-based applications. This approach considers the characteristics of DNNs partitioning and layers offloading over the cloud, edge and end devices.

The main contributions of this study are summarized as follows.

Xing Chen is the corresponding author.
This work is partly supported by the National Key R&D Program of China under Grant No.2018YFB1004800, the Natural Science Foundation of China under Grant No. 41801324 and No. 61672159, and the Natural Science Foundation of Fujian Province under Grant No. 2018J01619, No. 2019J01244 and No. 2019J01286.
B. Lin is with College of Physics and Energy, Fujian Normal University, Fujian Provincial Key Laboratory of Quantum Manipulation and New Energy Materials, Fuzhou，350117, China. The Fujian Provincial Collaborative Innovation Center for Optoelectronic Semiconductors and Efficient Devices, Xiamen, 361005, China. E-mail: WheelLX@163.com.
Y. Huang, J. Zhang, J. Hu and X. Chen are with the College of Mathematics and Computer Science, Fuzhou University, Fuzhou, 350117, China. E-mail: fzuhyh@foxmail.com, zhangjs0512@163.com, jackinhu@qq.com, chenxing@fzu.edu.cn.
Jun Li is with the School of Electronic and Optical Engineering, Nanjing University of Science and Technology, Nanjing, Jiangsu, 210094, China. E-mail: jun.li@njust.edu.cn.

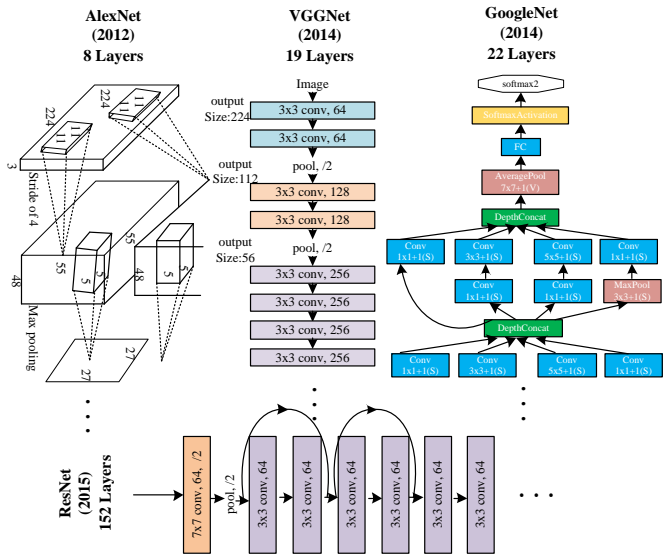

Fig. 1. Progression towards DNNs

1) A cost-driven offloading strategy based on PSO-GA is proposed to optimize the system cost during DNN layers offloading over the cloud, edge, and end devices.

2) A preprocessing operation is proposed to compress the number of layers in a DNN, which results in reducing the dimensions of a PSO particle.

3) PSO-GA is designed to enhance the population diversity of PSO and effectively reduce the sum cost of data transmission and layer computing while offloading DNN layers with deadline constraints.

The remainder of this paper is organized as follows. Section II reviews the related work. Section III discusses the process of cost-driven offloading for DNN-based applications in hybrid computing environments. Section IV presents the proposed PSO-GA algorithm in detail. Section V compares our algorithm with other state-of-the-art algorithms. Finally, section VI summarizes the full text and looks into future work.

## II. RELATED WORK

DNNs are popular in various fields such as natural language processing and computer vision. An offloading strategy is critical for the class of intelligent applications, which can rarely run on mobile devices due to the limited computation resources. Many work focused on offloading DNN layers from end devices to the cloud. Fang et al. [15] first designed a heuristic method to efficiently schedule heterogeneous servers for inference in DNNs, which satisfied the requirement on processing throughput and kept low response delay. Then they proposed a deep reinforcement learning (RL) method maximizing Quality of Service (QoS) from learning to schedule. This work optimized both response delay and inference accuracy, and ignored the data transmission between two layers in a DNN. Qi et al [16] designed a DNN-based object detection system with a model scheduling algorithm to adaptively offload DNN layers to the cloud based on the conditions of network and mobile devices. This work also aimed to reduce the system latency while deploying DNN-based applications. However, deploying DNNs only in the cloud will cause a prohibitively serious delay when offloading input sensor data to DNNs in the cloud.

MEC is a promising computing model for DNN-based applications by offloading DNN layers from resource-constrained mobile devices to the edge. Jeong [7] proposed a lightweight offloading system running on web-supported devices to offload DNN computations to edge servers. He designed a DNN partitioning algorithm to efficiently utilize the edge resources and reduce the system response time. However, this work ignored the difference in the computing capacity of each edge server. Lo et al. [10] made a dynamic DNN design for efficient workload allocation in edge computing. They explored the use of dynamic network structure and authentic operation (AO) unit to enhance DNNs, which had a better performance in reducing the amount of data transmissions under the same accuracy requirement. Jeong et al. [9] designed a simple method to offload DNN computations in the context of web applications. This approach could offload any DNN-based application to any edge server equipped with a browser. However, this work ignored how to offload each layer in DNN to a specific edge server for an optimized result.

It is a promising way to partition DNNs in hybrid computing environments, and combine the larger layers in the cloud and the smaller layers at the edge and end devices [5, 17]. Hu et al. [8] proposed a dynamic adaptive DNN surgery scheme that allowed DNN layers to be executed in both the cloud and the edge while limiting the data transmission under different network condition. This work focused on the DNN partitioning, and payed less attention to the DNN layers offloading. Teerapittayanon et al. [5] proposed distributed DNNs over distributed computing hierarchies, consisting of the cloud, the edge and end devices. This distributed DNNs exploited geographical diversity of sensors to reduce communication cost and improve object recognition accuracy. The environment in this work was similar to the one in this paper, but it did not focus on DNN layers offloading. Mao et al. [11] employed the differentially private strategies to enable the privacy preserving edge based training of DNN face recognition models, and the method is rigorously proved to be privacy preserving. Kang et al. [17] first examined DNN partitioning strategies that effectively offloaded DNN layers on the mobile devices and in the cloud to achieve low energy consumption and latency. Then they proposed a lightweight scheduler called Neurosurgeon to automatically partition DNNs between data-centers and mobile devices at the granularity of neural network layers. The scheduler had certain guiding significance for our research, however, it focused on DNNs partitioning instead of DNN layers offloading.

The work discussed above mostly focused on reducing the system response delay with the help of MEC [7, 10, 17]. There is little work aiming to minimize the system cost. The communication cost was considered in [5] while deploying distributed DNNs over distributed computing hierarchies. However, it ignored the layer computing cost. A DNN and a scientific workflow have many similarities, such as the overall structure, and the data dependencies between each pair of computing nodes. We have previously proposed a scheduling strategy for a deadline-constrained scientific workflow across multiple clouds [12]. The strategy aimed to minimize the execution cost of a scientific workflow within its deadline, which introduced the discrete PSO technique. This work has specific guiding significance for the work in this paper. However, it did not consider the affect of MEC. Cui et al. [18] proposed a data placement strategy based on GA for a scientific workflow to reduce the amount of data movement in cloud environment. They modified the mutation and crossover operator of GA to get a good performance from a global perspective.

In summary, previous studies have widely investigated the offloading strategies for DNN-based applications. However, it is still an open issue to optimize the system cost caused by data transmission and layer execution, while offloading DNN layers within the corresponding deadlines in hybrid computing environments.

## III. PROBLEM DEFINITION AND ANALYSIS

The purpose of our work is to minimize the system cost caused by data transmission and layer execution during DNN layers offloading while satisfying each DNN-based application's deadline constraint.

### A. Problem Definition

The problem definition includes the hybrid computing environments, some DNN-based applications, and an offloading strategy.

The hybrid computing environments $C = \{C_{cld}, C_{edg}, C_{dev}\}$ consist of the cloud, edge and end devices. The cloud $C_{cld} = \{s_1, s_2, ..., s_n\}$ consists of $n$ servers[1] in a region, and we only consider the offloading process in one region. The edge $C_{edg} = \{s_1, s_2, ..., s_m\}$ consists of $m$ servers in $m$ different regions, i.e., each region has only one server. There are $r$ IoT devices $C_{dev} = \{s_1, s_2, ..., s_r\}$, and each device has only one server. This study pursues an offloading scheme, therefore we focus on the computing power of each server and ignore their storage capacity. A server $s_i$ is expressed as

$$s_i = < p_i, c_i^{com}, t_i >, \quad (1)$$

where $p_i$ represents the measured computing power of server $s_i$, which is usually characterized by its CPUs [19]. $c_i^{com}$ represents the computation cost of $s_i$ per second[2], which is in a positive proportional relationship with $p_i$. $t_i = \{0, 1, 2\}$ represents the environment the server $s_i$ belongs to. When $t_i = 0$, $s_i$ belongs to the cloud, and it has strong computing power. When $t_i = 1$, $s_i$ belongs to the edge, and it has general computing power. When $t_i = 2$, $s_i$ belongs to the end devices, and it has poor computing power. The computing power of each server is assumed to be known and not fluctuate.

Formula (2) represents the bandwidth across different servers.

$$B = \begin{bmatrix} b_{1,1} & b_{1,2} & \cdots & b_{1,|C|} \\ b_{2,1} & b_{2,2} & \cdots & b_{2,|C|} \\ \vdots & \vdots & \cdots & \vdots \\ b_{|C|,1} & b_{|C|,2} & \cdots & b_{|C|,|C|} \end{bmatrix}, \quad (2)$$

$$b_{ij} = < \ell_{i,j}, c_{i,j}^{tran}, t_i, t_j >, \quad (3)$$

where $b_{i,j}$ is the bandwidth between server $s_i$ and server $s_j$. $\ell_{i,j}$ represents the value of bandwidth $b_{i,j}$, where $\forall i, j = 1, 2, ..., |C|$ and $i \neq j$. We do not consider ad hoc network [20], therefore there is no internet connection between two end devices (i.e., when $t_i$ and $t_j$ are both equal to 2, $\ell_{i,j}$ is 0). $c_{i,j}^{tran}$ represents the transmission cost per MB from server $s_i$ to server $s_j$. In addition, end devices connect to the edge via WIFI [21], therefore the servers belong to IoT devices only communicate with the servers belong to the edge within a certain range of WIFI radiation. The bandwidth is assumed to be known and not fluctuate.

The are many DNNs $G = \{G_1, G_2, ..., G_q\}$ from different end devices. A DNN is modeled as a directed acyclic graph $G_i = (L_i, E_i, D_i)$ [22], where $L_i = \{l_i^1, l_i^2, ..., l_i^s\}$ represents a finite set of nodes containing $s$ layers in $G_i$, $E_i = \{e_i^{1,2}, e_i^{1,3}, ..., e_i^{j,k}\}$ represents the data dependencies between each pair of layers, and $D_i = \{d_i^1, d_i^2, ..., d_i^n\}$ represents all the datasets including input data, intermediate data and output data in the $G_i$. Each DNN has a corresponding deadline $D(G_i)$, and an offloading strategy is named feasible solution if the DNN is completed within its deadline.

$e_i^{j,k} = (l_i^j, l_i^k)$ denotes a data dependency between layer $l_i^j$ and layer $l_i^k$, where layer $l_i^k$ is the direct successor of layer $l_i^j$, and layer $l_i^j$ is the direct precursor of layer $l_i^k$. In the process of offloading DNN layers, a layer cannot start executing until all of its precursors have been completed.

For a layer $l_i^j = < a_i^j, i_i^j, o_i^j >$, $a_i^j$ is the calculated amount of $l_i^j$, $i_i^j$ is the input datasets of $l_i^j$, and $o_i^j$ is the output datasets of $l_i^j$. Serial processing model [23] is adopted in the execution process, which means that a server can only execute one layer at the same time and a whole layer is executed on a same server. Therefore, the execution time $T_{exe}(l_i^j, s_k)$ that offloading layer $l_i^j$ to a ready server $s_k$ is calculated as (4). The layers and datasets are many-to-many correspondence (i.e., a layer may require many input datasets from different servers, and a dataset may be used by many layers) shown as GoogleNet in Fig. 1.

$$T_{exe}(l_i^j, s_k) = \frac{a_i^j}{p_k}. \quad (4)$$

For a dataset $d_i^j = < \partial_i^j, g_i^j, c_i^j, \Omega_i^j, f_i^j >$, $\partial_i^j$ is the dataset size, $g_i^j$ is the layer generating $d_i^j$ using (5), where $D_{ini}$ is the initial datasets (i.e., the input datasets of a DNN), $D_{gen}$ is the generated datasets (i.e., the intermediate datasets generated during DNN execution) and $L(d_i^j)$ is the layer generating dataset $d_i^j$. In addition, $c_i^j$ is the layer consuming $d_i^j$, $\Omega_i^j$ is the original server storing $d_i^j$, and $f_i^j$ is the final server using $d_i^j$.

---

[1] In order to have a unified expression for the computing resource in different platforms, we use 'server' instead of 'virtual machine' to express the instances in cloud.

[2] Although the computation cost in cloud is measured in hour, the execution time of DNN is usually at the millisecond level. Therefore, we use 'second' instead of 'hour' to measure the computation cost of each server.

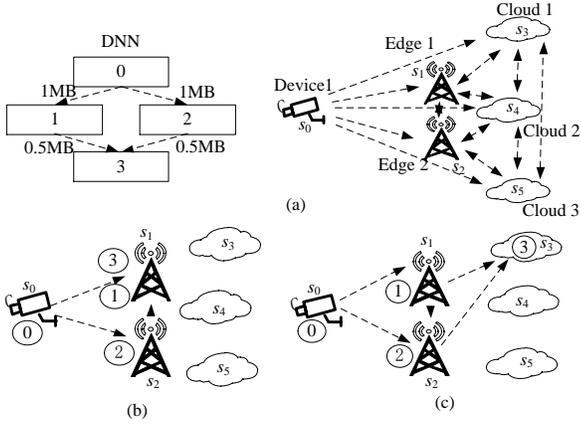

Fig. 2. A sample of cost-driven offloading for a DNN.

Table I. The execution time of each layer on different servers

|       | $s_0$ | $s_1$ | $s_2$ | $s_3$ | $s_4$ | $s_5$ |
|-------|-------|-------|-------|-------|-------|-------|
| $l_0$ | 1.1s  | -     | -     | -     | -     | -     |
| $l_1$ | 1.92s | 0.98s | 0.62s | 0.31s | 0.19s | 0.09s |
| $l_2$ | 2.35s | 1.20s | 0.75s | 0.67s | 0.41s | 0.32s |
| $l_3$ | 2.12s | 1s    | 0.8s  | 0.56s | 0.45s | 0.21s |

Table II. The cost of six servers

| Servers | Cost/hour($) |
|---------|--------------|
| $s_0$   | 0            |
| $s_1$   | 10           |
| $s_2$   | 15           |
| $s_3$   | 1            |
| $s_4$   | 2            |
| $s_5$   | 3            |

Table III. The bandwidth between two categories and the corresponding cost

| $t_i$ |   | $t_j$ | $\ell_{i,j}$ (M/s) | $c_{i,j}^{tran}$ ($/GB) |
|-------|---|-------|--------------------|-------------------------|
| 0     | ←→ | 0    | 5                  | 0.4                     |
| 0     | ←→ | 1    | 2                  | 0.8                     |
| 0     | ←→ | 2    | 2                  | 0.8                     |
| 1     | ←→ | 1    | 10                 | 0.16                    |
| 1     | ←→ | 2    | 10                 | 0.16                    |

Therefore, the transmission time $T_{\text{trans}}(d_i^j, s_k, s_r)$ that transferring dataset $d_i^j$ from server $s_k$ to server $s_r$ is calculated as (6).

$$g_i^j = \begin{cases} 0, & d_i^j \in D_{\text{ini}} \\ L(d_i^j), & d_i^j \in D_{\text{gen}} \end{cases}, \quad (5)$$

$$T_{\text{trans}}(d_i^j, s_k, s_r) = \frac{\partial_i^j}{\ell_{k,r}}. \quad (6)$$

The purpose of the offloading strategy is to minimize the system cost caused by data transmission and layer execution, with the deadline constraints of all DNN-based applications. Any layer execution in a DNN have to satisfy both conditions: (1) the layer has been offloaded to a specific server; (2) the datasets required by this layer have been transferred to the same server. The offloading strategy for DNN-based applications is defined as $O = (C, L_i, D_i, M, T_i^{comp}, C_{\text{total}})$, where $M = \bigcup_{i=1,2,...,|C|} \{<l_i^j, s_s> \cup <d_i^j, s_k, s_r>\}$ represents the map set from any DNN layers $L_i$ and datasets $D_i$ to hybrid computing environments $C$. $<l_i^j, s_s>$ represents that the layer $l_i^j$ is offloaded to server $s_s$, and $<d_i^j, s_k, s_r>$ represents that dataset $d_i^j$ is transferred from server $s_k$ to server $s_r$. If all $<l_i^j, s_s>$ sets are determined, then all $<d_i^j, s_k, s_r>$ sets are determined. Therefore, the map can be modified as $M = \bigcup_{i=1,2,...,|C|} \{<l_i^j, s_s>\}$. $T_i^{comp}$ represents the completion time of DNN $G_i$, and $C_{\text{total}}$ represents the whole system cost for executing all DNN layers.

$$T_i^{comp} = \max_{l_i^j \in L_i} \{T_{\text{comp}}(l_i^j)\}, \quad (7)$$

$$C_{\text{total}} = \sum_{i=1}^{|C|} c_i^{com} \cdot (T_{\text{off}}(s_i) - T_{\text{on}}(s_i)) \\ + \sum_{j=1, \Omega_j^m=j}^{|C|} \sum_{k=j+1, f_j^m=k}^{|C|} c_{j,k}^{tran} \cdot \partial_i^j, \quad (8)$$

where $T_{\text{comp}}(l_i^j)$ is the completion time of layer $l_i^j$, $T_{\text{off}}(s_i)$ is the turn-off time of server $s_i$, and $T_{\text{on}}(s_i)$ is the turn-on time of server $s_i$. Assuming that when the first task arrives on a server, the server is turned on immediately with no delay, and when the last task on a server is completed, the server is turned off immediately with no delay.

The problem of the cost-driven offloading for DNN-based applications over the cloud, edge, and end devices can be formalized as (7). Its core purpose is to pursue a minimum total system cost while satisfying the deadline constraint for each DNN-based application.

$$\begin{aligned} & \text{Minimize } C_{\text{total}} \\ & \text{subject to } \forall i, T_i^{comp} \leq D(G_i) \end{aligned}. \quad (9)$$

*B. Problem Analysis*

Fig. 2(a) is a sample of offloading for a DNN $G_i$, which includes four layers $\{l_i^0, l_i^1, l_i^2, l_i^3\}$, and four transmission datasets $\{d_i^1, d_i^2, d_i^3, d_i^4\}$, whose size are {1MB, 1MB, 0.5MB, 0.5MB}. Its deadline is 3.7s. Note that layer $l_i^0$ must be executed on the end device (i.e., $s_0$). There are six servers in hybrid computing environments. Table I shows the execution time of each layer on the six servers. Table II shows the cost of six servers. Table III shows the bandwidth between two categories and the corresponding cost.

Fig. 2(b) is the offloading result according to a greedy strategy [24]. The completion time of the DNN is 3.65s, and the system cost is 0.0044013 with greedy strategy. Fig. 2(c) is the optimal offloading result. The completion time of the DNN is 3.41s, and the system cost is 0.0036517. The system cost with the optimal strategy is 18.18% less than the former. The greedy offloading strategy offloads each layer to the server with the lowest cost step by step within the corresponding deadline. However, each step with the best choice for offloading a layer from a local perspective does not necessarily get the best result. In this study, the offloading issue for many DNNs have to be addressed. We need a suitable approach to offload all DNN layers from a global perspective.

IV. OFFLOADING STRATEGY BASED ON PSO-GA

For an offloading strategy $O = (C, L_i, D_i, M, T_i^{comp}, C_{\text{total}})$, its core purpose is to find a map from all $L_i$ to $C$ that has minimum system total cost $C_{\text{total}}$ while each DNN completion time $T_i^{comp}$ is not more than their corresponding deadline $D(G_i)$. Finding the best map from all $L_i$ to $C$ is an NP-hard problem [25]. Therefore, an offloading strategy based on PSO-GA algorithm to optimize the system total cost is proposed from a global perspective in hybrid computing environments. To improve the efficiency of our offloading strategy, a preprocessing for a DNN is designed to compress the amount of layers and data transmission. This section mainly describes the preprocessing for a DNN, and PSO-GA algorithm as follows.

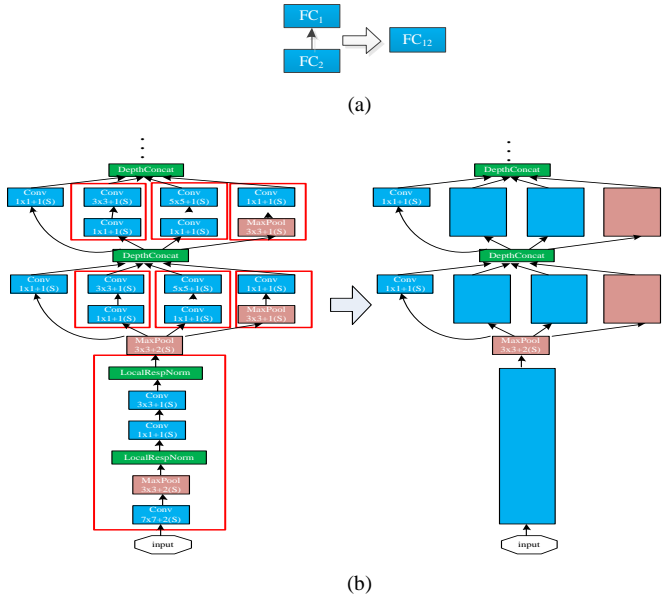

Fig. 3 Preprocessing for a DNN: (a) Merging two adjacent layers into a new one; (b) The structure of GoogleNet before and after preprocessing

*A. Preprocessing for a DNN*

Algorithm 1 describes the preprocessing pseudocode for a DNN that merges two adjacent layers with a common cut-edge into a new layer. Two adjacent layers are merged into a new layer when the precursor's out-degree and the successor's in-degree are both 1. A cut-edge is the data dependence between the precursor and successor, which disappears after preprocessing. The procedure of merging two adjacent layers with a common cut-edge into a new layer is shown in Fig. 3(a). GoogleNet [26] has many cut-edge layers, and the number of compressed layer reaches about 48%. Fig. 3(b) depicts the structure of GoogleNet before and after preprocessing. PSO-GA has better efficiency for a DNN with less layers.

**Property 1**: Preprocessing reduces the number of layers in a DNN, and improves the efficiency of PSO-GA execution. However, it may influence the final offloading result.

Fig. 3 depicts that the number of layers in GoogleNet is reduced. Part B in this section describes that the particle encoding dimension of PSO-GA depends on the number of all layers. The efficiency of PSO-GA execution is mainly based on the particle encoding dimension. Therefore, the decrease of layers will improve the efficiency of PSO-GA execution. In Fig. 3(a), $FC_1$ and $FC_2$ are merged into $FC_{12}$, which means that $FC_1$ and $FC_2$ have to be offloaded to a server together. However, $FC_1$ and $FC_2$ can be offloaded to two different servers if they are not merged together. Therefore, preprocessing may influence the final offloading result.

**Algorithm 1**: Merge two adjacent layers into a new layer
**procedure** preprocess (*G*(*L*, *E*, *D*))
  1: Record in-degree and out-degree of all layers in *G*
  2: If there are two adjacent layers with a common cut-edge, merging these two layers into a new layer.
  3: Repeat *step 2* until there is no cut-edge.
**end procedure**

| layers | 0 | 1 | 2 | 3 |
|---|---|---|---|---|
| offloading location | 0 | 1 | 2 | 3 |

Fig. 4 An encoded particle corresponding to the offloading for a DNN

*B. PSO-GA*

PSO algorithm is an evolutionary computation technique, which is inspired by social behavior observed in nature, such as bird flocks. In 1995, it was first introduced by Eberhart and Kennedy [27]. The particle is an important concept in PSO, whose position represents a candidate solution to the optimization problem. Each particle searches for better position by modifying its own velocity, which determines its future magnitude and direction. The velocity and current position determine the movement of a corresponding particle. and they are iteratively updated by (10) and (11).

$$V_i^{t+1} = w \times V_i^t + c_1 r_1 (pBest_i^t - X_i^t) + c_2 r_2 (gBest^t - X_i^t), \quad (10)$$

$$X_i^{t+1} = X_i^t + V_i^{t+1}. \quad (11)$$

where $V_i^t$ and $X_i^t$ represent the velocity and position of the $i^{th}$ particle at the $t^{th}$ iteration, respectively. From (10), you can find that each particle's velocity is influenced by its personal best position, *pBest*, and the global best position of its population, *gBest*. Inertia weight $w$ determines the extent to which the previous velocity affects the current velocity, which impacts on the convergence of PSO. $c_1$ and $c_2$ represent a particle's cognitive ability to its personal best value and global best value, which are both acceleration coefficients. $r_1$ and $r_2$ are introduced to enhance the random search for an optimization result, whose value are both between 0 and 1.

PSO has been widely used to solve continuous optimization problems. Offloading DNN layers in hybrid computing environments is a discrete problem, and it needs a new coding approach. In addition, a suitable strategy for particle update should be introduced to avoid the premature convergence of traditional PSO. In this paper, PSO-GA is proposed to solve the above shortages of traditional PSO. The offloading strategy based on PSO-GA for DNNs is described as follows.

**1) Problem Encoding**

A good encoding strategy can enhance PSO-based algorithm's search efficiency and improve its performance, which usually satisfies the following three principles [28].

**Definition 1** (**Completeness**). Each candidate solution can be encoded as a particle in the problem space.

**Definition 2** (**Non-redundancy**). Each candidate solution has only one corresponding encoded particle in the problem space.

**Definition 3** (**Viability**). Each encoded particle represents a candidate solution in the problem space.

Designing an encoding strategy that simultaneously satisfies the three principles is difficult. Inspired by [29], we adopt a server-order nesting strategy to encode the layers offloading problem. A particle represents a candidate solution of cost-driven offloading for all DNNs in hybrid computing environments, and the $i^{th}$ particle in the $t^{th}$ iteration is described in (12).

$$X_i^t = (x_{i1}^t, x_{i2}^t, \ldots, x_{ip}^t), \quad (12)$$

$$x_{ik}^t = (s_j, \varphi_j)_{ik}^t, \quad (13)$$

where $p$ is the total number of layers from all ready DNN-based applications. $x_{ik}^t$ ($k=1, 2, \ldots, p$) indicates the final offloading location of the $k^{th}$ layer in the $t^{th}$ iteration in (13). It means that the $k^{th}$ layer is offloaded to server $s_j$ with a specified order $\varphi_j$, whose value ranges from 0 to $p$-1. The order of each layer in the same particle is different from each other, and the layer with smaller order is processed earlier when there are more than two layers on the same server. Therefore, the particle dimension is twice of the total number of layers. Fig. 4 shows an encoded particle corresponding to the offloading for a DNN in Fig. 2(c).

**Property 2**: The encoding strategy meets completeness and non-redundancy principles, but it may not meet the viability principle.

After DNN layers offloading, each layer is offloaded to the corresponding server with a specified order. A layer can be offloaded to any server with any order under the data dependence constraint, and the dimension in a particle can be the corresponding value of server and order. Therefore, each offloading strategy has the corresponding particle, which meets the completeness principle. An offloading strategy for DNNs corresponds to a 2$p$-dimensional particle. The value of the $k^{th}$ dimension in a particle are the server number processing the $k^{th}$ layer and the specified process order. Therefore, an offloading strategy only corresponds to a specified particle, which meets the non-redundancy principle. However, some candidate solutions corresponding to the particles may not meet the deadline constraints. For example, if the final offloading location of layers in Fig. 2 is (0, 0, 2, 3), then layer $l_0$ and layer $l_1$ are offloaded to end device $D_1$. The completion time of this DNN is more than 4s, which exceeds its deadline (that is, 3.7s). Therefore, the encoding strategy may not meet the viability principle.

**2) Fitness Function**

The fitness function is used to evaluate the performance of all particles. In general, a particle with smaller fitness represents a better candidate solution [30]. This work pursues minimum system cost for offloading DNN layers while satisfying their deadline constraints. Therefore, a particle with lesser system cost can be considered as a better solution. However, the encoding strategy designed in this paper fails to meet the viability principle.

All particles can be divided into two categories: feasible particles and infeasible particles, whose definitions in this paper are described as follows.

**Definition 4 (Feasible particle)**. A particle that corresponds to a DNN layers offloading strategy meets all deadline constraints.

**Definition 5 (Infeasible particle)**. A particle that corresponds to a DNN layers offloading strategy fails to meet all deadline constraints (*i.e.*, At least one DNN's completion time exceeds its corresponding deadline).

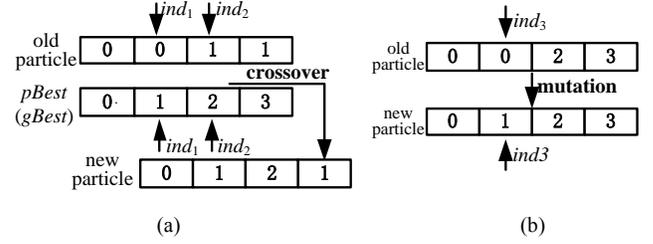

Fig. 5 Update operation: (a) Crossover operator for *individual* (*social*) *cognition* component; (b) Mutation operator for *inertia* component

The definition of fitness function that compares two candidate solutions has to be modified according to three different situations.

**Case 1**: Both particles are feasible. The particle with lesser system cost is selected, and the fitness function is defined as follows.

$$F(X_i) = C_{\text{total}(X_i)}. \quad (14)$$

**Case 3**: One particle is feasible, and the other one is infeasible. The feasible particle is selected, and the fitness function is defined as follows.

$$F(X_i) = \begin{cases} 0, \text{if } \forall i, T_i^{comp}{}_{(X_i)} \leq D(G_i) \\ 1, \text{else} \end{cases}. \quad (15)$$

**Case 2**: Both particles are infeasible. The particle with lesser total completion time is selected, and this particle is more likely to become a feasible particle after update operations. The fitness function is defined as follows.

$$F(X_i) = \sum T_i^{comp}{}_{(X_i)}. \quad (16)$$

**3) Update Strategy**

Formula (10) shows that PSO has three main parts: *inertia*, *individual cognition*, and *social cognition*. The iterative update of each particle is affected by both its personal best positon and the global best position in current generation [31]. Prematurely falling into a local optimum is a major defect of traditional PSO. To enhance the search ability of our algorithm and avoid prematurely convergence, the crossover operator and mutation operator of GA are introduced for particle update. The iterative update of the $i^{th}$ particle at the $t^{th}$ iteration is shown as (17), where $C_g()$ and $C_p()$ represent crossover operators, and $M_u()$ is mutation operator. In addition, the value of the order $\varphi_j$ for each layer remains the same, and only the value of the server $s_j$ for each layer is updated.

$$X_i^t = c_2 \oplus C_g(c_1 \oplus C_p(w \oplus M_u(X_i^{t-1}), pBest_i^{t-1}), gBest^{t-1}), \quad (17)$$

For *individual cognition* component and *social cognition* component, the crossover operator of GA is introduced to refresh the corresponding component of (10), which is describe as (18) and (19), respectively.

$$B_i^t = c_1 \oplus C_p(A_i^t, pBest^{t-1}) = \begin{cases} C_p(A_i^t, pBest^{t-1}) & r_1 < c_1 \\ A_i^t & else \end{cases}, \quad (18)$$

$$C_i^t = c_2 \oplus C_g(B_i^t, gBest^{t-1}) = \begin{cases} C_g(B_i^t, gBest^{t-1}) & r_2 < c_2 \\ B_i^t & else \end{cases}, \quad (19)$$

where $r_1$ and $r_2$ are both random factors, whose value is between 0 and 1. $C_p()$ (or $C_g()$) is crossover operators, which randomly chooses two locations in a particle to be updated, and then replaces its segment between the two locations with the segment in the same interval in *pBest* (or *gBest*) particle. The crossover operator for *individual* (or *social*) *cognition* component is shown as Fig. 5(a). It randomly chooses $ind_1$ and $ind_2$ locations in an *old* particle, and then replaces the segment between $ind_1$ and $ind_2$ with the *pBest* (or *gBest*) particle in the same interval.

**Property 3**: A particle can change from infeasible to feasible after crossover operator, and vice versa.

The encoded particle (0, 0, 1, 1) in Fig. 5(a) is infeasible, whose completion time is more than 5s in Fig. 2. The *pBest* particle is (0, 1, 2, 3), and the crossover locations are 2$^{nd}$ and 3$^{rd}$. Therefore, the generated encoded particle is (0, 1, 2, 1) after the crossover operator. The completion time of the new particle is 3.65s, which is less than the corresponding deadline. This generated particle is feasible. On the contrary, a feasible particle (0, 3, 3, 3) crossover with the *pBest* particle (0, 1, 2, 3) in location 1$^{st}$ and 2$^{nd}$. The new generated particle (0, 1, 3, 3) is infeasible.

For *inertia* component, the mutation operator of GA is introduced to refresh the corresponding component of (10), which is described as (20).

$$A_i^t = w \oplus M_u(X_i^{t-1}) = \begin{cases} M_u(X_i^{t-1}) & r_3 < w \\ X_i^{t-1} & else \end{cases}, \quad (20)$$

where $r_3$ is a random factor, whose value is between 0 and 1. $M_u()$ is mutation operator, which randomly chooses a location in a particle, and then alters the value of the server $s_j$ in the range of $|C|$. Fig. 5(b) illustrates the mutation operator for *inertia* component. It randomly chooses $ind_3$ location, and then alters the value of $ind_3$ from 0 to 1.

**Property 4**: A particle can change from infeasible to feasible after mutation operator, and vice versa.

The mutation operator randomly selects 2$^{nd}$ location of an infeasible particle (0, 0, 2, 3) to mutate, and then generates a new feasible particle (0, 1, 2, 3). This particle is the optimal solution in Fig. 2. Alternately, it mutates a feasible particle (0, 1, 2, 3) in location 2$^{nd}$, and then generates a new infeasible particle (0, 0, 2, 3). The completion time of this particle is more than 4s, which exceeds its deadline (that is, 3.7s).

#### 4) Map from a Particle to DNN Layers Offloading

Algorithm 2 shows the pseudocode of mapping a particle to cost-driven offloading for DNNs in hybrid computing environments. The inputs include DNNs $G$, resource of the hybrid computing environments $C$ and a particle $X$. First, the whole data transmission cost $C_{trans}$ and the whole system cost $C_{total}$ are set to 0 (line 1). After initialization, the layers are offloaded to the corresponding servers. According to the layer execution sequence, the layer $l_i^j$ is offloaded to server $s_{x(j)}$. If $l_i^j$ has no parents (i.e., $l_i^j$ is input layer), its start time $T_{start}(l_i^j)$ is equal to the leased time $T_{lease}(s_{x(j)})$ of server $s_{x(j)}$. Otherwise, the start time of layer $l_i^j$ has to consider the data transmission time from its parents (line 3-12). Obviously, the end time $T_{end}(l_i^j)$ of layer $l_i^j$ is equal to the sum of its start time and execution time on server $s_{x(j)}$ (line 13-14). In addition, if the end time of any layer exceeds its corresponding deadline, this particle is infeasible and the algorithm execution is terminated (line 15-17). The update of the $s_{x(j)}$ leased time $T_{lease}(s_{x(j)})$ has to consider the data transmission

---

**Algorithm 2**: A map from a particle to DNN layers offloading
**procedure** DNNsOffloading(*G*, *C*, *X*)
1: Initialization: $C_{trans} \leftarrow 0$, $C_{total} \leftarrow 0$
2: **for** $j = 0$ to $j = |G|-1$
3:     **if** $l_i^j$ has no parents **then**
4:         $T_{start}(l_i^j) = T_{lease}(s_{x(j)})$
5:     **else**
6:         maxTrans = 0
7:         **foreach** parent $l_i^p$ of $l_i^j$ **do**
8:             maxTrans = **max** (maxTrans, $\partial_i^p / \ell_{x(p),x(j)}$)
9:             $C_{trans} \mathrel{+}= c_{x(p),x(j)}^{tran} \cdot \partial_i^p$
10:         **end for**
11:         $T_{start}(l_i^j) = T_{lease}(s_{x(j)}) +$ maxTrans
12:     **end if**
13:     $exe = T_{exe}(l_i^j, s_{x(j)})$
14:     $T_{end}(l_i^j) = T_{start}(l_i^j) + exe$
15:     **if** $T_{end}(l_i^j) > D(G_i)$ **then**
16:         **return** this particle is infeasible
17:     **end if**
18:     *transfer* = 0
19:     **foreach** child $l_i^c$ of $l_i^j$ **do**
20:         *transfer* $\mathrel{+}= \partial_i^j / \ell_{x(j),x(c)}$
21:     **end for**
22:     $T_{lease}(s_{x(j)}) = T_{lease}(s_{x(j)}) + exe + transfer$
23:     **if** $l_i^j$ is output layer **then**
24:         $T_i^{comp} = T_{end}(l_i^j)$
25:     **end if**
26: calculate $C_{total}$ according to formula (8)
**end procedure**

---

from layer $l_i^j$ to its children layers (line 18-22). If $l_i^j$ is output layer, the completion time of DNN $G_i$ is equal to the end time of layer $l_i^j$ (line 23-25). Finally, the whole system cost $C_{total}$ is calculated according to formula (8).

#### 5) Parameter Settings

The inertia weight $w$ in (10) affects the convergence and search ability of PSO [32]. Formula (21) represents an adjustment mechanism for the inertia weight [33].

$$w = w_{max} - iters_{cur} \times \frac{w_{max} - w_{min}}{iters_{max}}. \quad (21)$$

where $w_{min}$ and $w_{max}$ are the given minimum and maximum of $w$ in the initialization phase. $iters_{cur}$ and $iters_{max}$ are the current number and the maximum number of iterations, respectively. Therefore, the algorithm focuses on the global search at the beginning of execution. With the increase of iterations, the value of $w$ reduces linearly and the algorithm gradually focuses on the local search.

The adjustment mechanism in (21) fails to meet the nonlinear characteristics of DNN layers offloading. Therefore, this paper designs an adjustment mechanism that can adaptively adjust the search ability according to the quality of current particle in (22).

$$w = w_{max} - (w_{max} - w_{min}) \times \exp(d(X_i^{t-1}) / (d(X_i^{t-1}) - 1.01)), \quad (22)$$

$$d(X^{t-1}) = \frac{div(gBest^{t-1}, X^{t-1})}{|C|}, \quad (23)$$

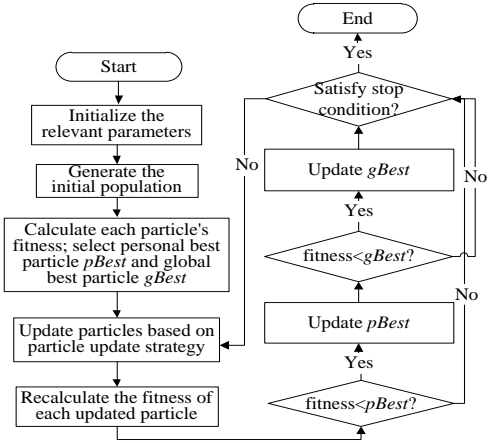

Fig. 6 The flowchart of PSO-GA

where $div(gBest^{t-1}, X^{t-1})$ indicates the number of difference in $s_j$ coordinate between the global best particle $gBest^{t-1}$ and the current particle $X^{t-1}$. This mechanism can adaptively adjust its search ability according to the difference between the global best particles and current particle. When $div(gBest^{t-1}, X^{t-1})$ is small, it means that there is a small gap between $gBest^{t-1}$ and $X^{t-1}$, and the algorithm prefer to enhance the local search and accelerate the convergence to find an optimal solution.

The other two acceleration coefficients (*i.e.*, $c_1$ and $c_2$) are given by [34]. Note that $c_1^{start}$ and $c_1^{end}$ are the start value and end value of $c_1$. $c_2^{start}$ and $c_2^{end}$ are the start value and end value of $c_2$.

**6) Algorithm Flowchart**

Fig. 6 is the flowchart of PSO-GA, whose detailed process is described as follows.

*Step* **1**: Initialize the relevant parameters of PSO-GA, such as population size $S_{pop}$, maximum iteration, inertia weight, and cognitive factors, and then generate the initial population.

*Step* **2**: According to the map from a particle to DNN layers offloading (Algorithm 2), calculate the fitness of each particle based on (14) to (16). Each particle is set as its personal best particle, and the particle with the smallest fitness is set as the global best particle in current generation.

*Step* **3**: Update particles based on (17) to (20), and recalculate the fitness of each updated particle.

*Step* **4**: If the fitness of the updated particle is smaller than its personal best particle, set the updated particle as its own personal best particle. Otherwise, go to *Step* **6**.

*Step* **5**: If the fitness of the updated particle is smaller than the global best particle, set the updated particle as the global best particle.

*Step* **6**: Verify whether the stop condition is met. If it is not satisfied, go to *Step* **3**. Otherwise, terminate the procedure.

## V. EXPERIMENTAL RESULTS AND ANALYSIS

All the simulation experiments were conducted on a Win8 64-bit operating system with a G3250 3.20 GHz Intel (R) Pentium (R) processor and 8GB RAM. The corresponding parameters of PSO-GA were set based on [33]: $S_{pop}$ = 100, $iters_{max}$ = 1000, $w_{max}$ = 0.9, $w_{min}$ = 0.4, $c_1^{start}$ = 0.9, $c_1^{end}$ = 0.2, $c_2^{start}$ = 0.4, and $c_2^{end}$ = 0.9.

[3] https://github.com/LinBin403/dataset-for-our-research

Table IV. The configurations and the cost for all servers

| Servers | Configurations | Cost/hour($) | $t_i$ |
|---|---|---|---|
| $\{s_1, s_2, ..., s_{10}\}$ | 2CPUs 4GB | 0 | 2 |
| $\{s_{10}, s_{11}, ..., s_{15}\}$ | 16CPUs 32GB | 2.43 | 1 |
| $s_{16}$ | 4CPUs 8GB | 0.225 | 0 |
| $s_{17}$ | 8CPUs 16GB | 0.45 | 0 |
| $s_{18}$ | 16CPUs 32GB | 0.9 | 0 |
| $s_{19}$ | 32CPUs 64GB | 1.8 | 0 |
| $s_{20}$ | 64CPUs 128GB | 3.6 | 0 |

### A. Experimental Setup

We conducted our experiments using four types of DNNs: AlexNet, VGG19, GoogleNet and ResNet101. The structure, datasets and computing amount in each type of DNN are different. The basic layer execution time, data transmission amount between two layers, and the overall structure for our tested DNNs is recorded in a file[3] in github.com.

The hybrid computing environments have 20 servers $\{s_1, s_2, ..., s_{20}\}$, which are divided into 3 categories. The first 10 servers belong to the end devices, the last 5 servers belong to the cloud, and the other 5 servers belong to the edge. The processing capacity of a server in the same category is roughly proportional to its cost. We assume that the end servers have lowest configurations and execute DNN layers without charging. Table IV shows the configurations and the cost for all servers. In this paper, each end server is connected to two nearby edge servers. To simplify the experiments, we calculate the bandwidths and the corresponding cost by 3 categories. The bandwidth between two categories and the corresponding transmission cost is shown in Table III.

Finally, each DNN needs a specific deadline to verify whether an offloading strategy is feasible or not. We set five different deadlines for each DNN as follows.

$$D_j(G_i) = r_j \cdot H(G_i), r_j = \{1.2, 1.5, 3, 5, 8\}, \quad (24)$$

where $H(G_i)$ is the execution time of a DNN $G_i$ based on HEFT algorithm [35].

### B. Competitive Algorithms

To verify the effectiveness of PSO-GA, we modified the GA [18] and the Greedy [24] to adapt the cost-driven offloading strategy for DNNs in hybrid computing environments.

GA adopts a binary encoding strategy, and its fitness function is according to Formula (14) - (16). The map from an encoded chromosome to a cost-driven offloading solution should not only consider the computation cost for each layer, but also the transmission cost for each dataset.

Greedy offloads each layer to the cheapest server within the corresponding deadline. If a layer offloaded to the cheapest server cannot meet the deadline constraint, then the layer has to be offloaded to the second cheapest server. It follows these operations and iterates over.

Finally, prePSO is selected as another comparison algorithm, which is the PSP-GA with preprocessing discussed in section IV.

### C. Experimental Results and Analysis

GA, PSO-GA, and prePSO belong to the meta-heuristic algorithms. They are terminated if they maintain the same value in 50 iterations. The offloading results may be different with the same configurations in each experiment. Therefore, the system cost is measured as the average value of 50 repeated experiments. If an infeasible solution occurs, its system cost is expressed as a negative value.

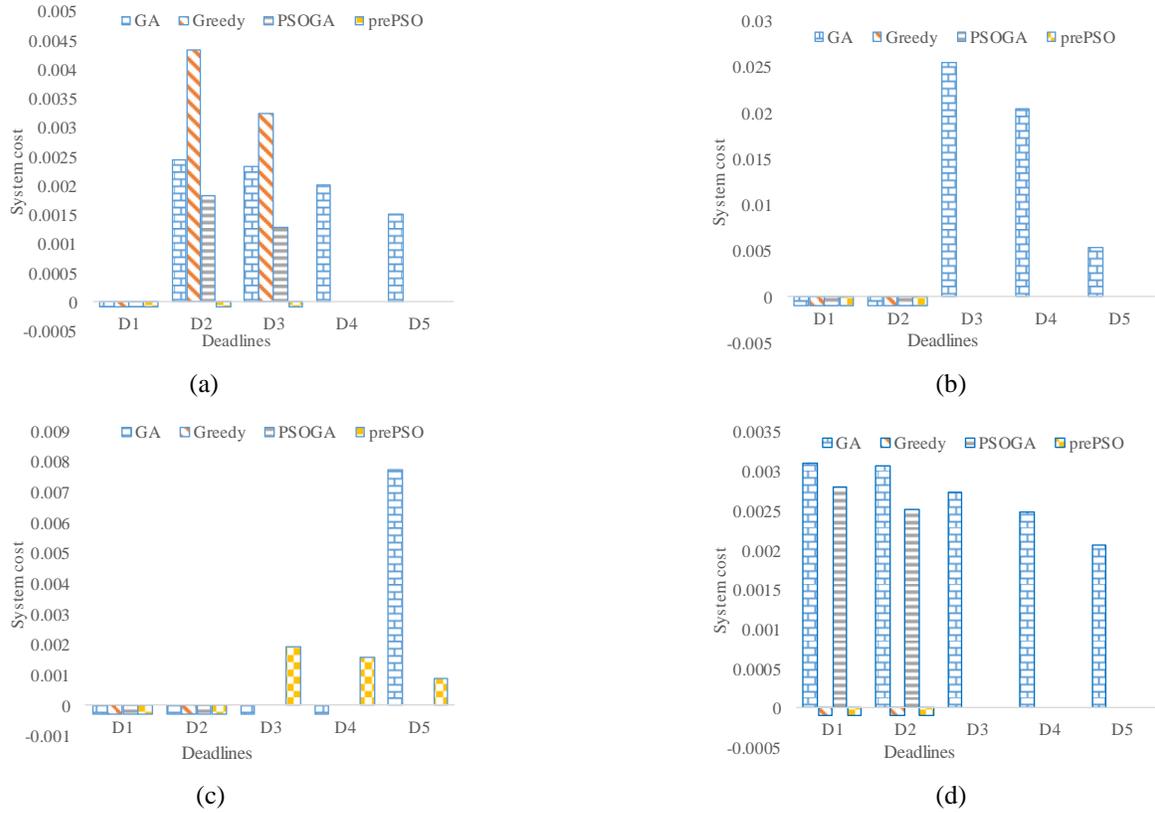

Fig. 7 The system cost of different strategies for one DNN per end device: (a) AlexNet; (b) VGG19; (c) GoogleNet ; (d) ResNet101

Fig. 7 shows The system cost of different strategies for one DNN per end device. In these experiment, there is only one DNN on each end device originally. It means that there are 10 DNNs on 10 end devices, respectively. In general, the system cost becomes less and less as the deadline is gradually loose for all offloading strategies. With the looser deadline constraints, more layers can be offloaded to the cheaper servers under the same situation. PSO-GA has the best performance due to that it evolves iteratively from a global perspective. Greedy is an extreme strategy. When the deadline is loose, it can find the feasible solution. On the contrary, it sometimes cannot find a feasible solution. The search scope of GA is relatively limited during each iteration, and it does not adaptively adjust according to the performance of the current chromosome. This results are worse compared with PSO-GA. For VGG19 and ResNet101, prePSO compresses all the layers into one layer in a DNN. Therefore, it offloads all layers of a DNN to its corresponding end devices, and its performance is similar to Greedy for VGG19 and ResNet101. For GoogleNet, there is a big gap between PSO-GA and prePSO after $D_3(G)$. The reason is that the preprocessing influences the final offloading result discussed as **Property 1**. The compressed layers have larger computation amount, which have to be offloaded to the servers with more computing power.

Fig. 7(a) shows The system cost of different strategies for one AlexNet per end device. The strategies in Fig. 7(a) have less system cost compared with those in Fig. 7(b) and Fig. 7(d).

This is mainly due to that the number of layers, the average amount of each dataset, and the average execution time of each layer in Fig. 7(a) are all much less than those in Fig. 7(b) and Fig. 7(d). This results in that the system cost among the three figures is not an order of magnitude. In Fig. 7(c), there is no strategy offering a feasible solution before after $D_3(G)$. Although the deadline is expanded by 1.5 times the completion time by HEFT algorithm for each DNN, each layer has to be executed according to the serial processing model on a specific server. The number of servers is limited, and the number of layers for ResNet101 is more than 1000 in hybrid computing environments.

Fig. 8 shows The system cost of different strategies for three DNNs per end device. It means that there are three DNNs on each end device originally. From Fig. 7, we find that almost all offloading strategies cannot achieve a feasible solution with $D_1(G)$ and $D_2(G)$ constraints. The number of layers in the experiments of Fig. 8 is three times that in Fig. 7. In order to avoid the generation of more infeasible solutions, the deadlines in the experiments of Fig. 8 is twice that in Fig. 7.

In general, the system cost in Fig. 8 is almost 4 times that in Fig. 7. It is obvious the system cost becomes less and less as the deadline is gradually loose for all offloading strategies. Greedy has the worst performance due to that it is an extreme strategy from a local perspective. It prioritizes scheduling each layer to the cheapest server within its DNN corresponding deadline step by step. As the total number of DNN layers increase, the higher layers usually fails to be completed with the deadline constraint by Greedy strategy. Through an overview of Fig. 8, we find that Greedy achieves feasible solutions only within much looser deadline (i.e., $D_5(G)$) for AlexNet and ResNet101. PSO-GA has the best performance. It averts the premature convergence of PSO and improves the diversity of population. This results in that a layer with a large amount of computation is offloaded to the cloud, and a layer with a large amount of data transmission is offloaded the edge and end devices according to the deadline constraints. The overall trend of different strategies for system cost in Fig. 8 is similar to that in Fig. 7. From the experiments, we find that data transmission cost accounts for the majority of the total system cost for GoogleNet and ResNet101. The reason for this result is that they have more layers, and the amount of layer computation is relatively smaller compared with AlexNet, whose number of layers is 11.

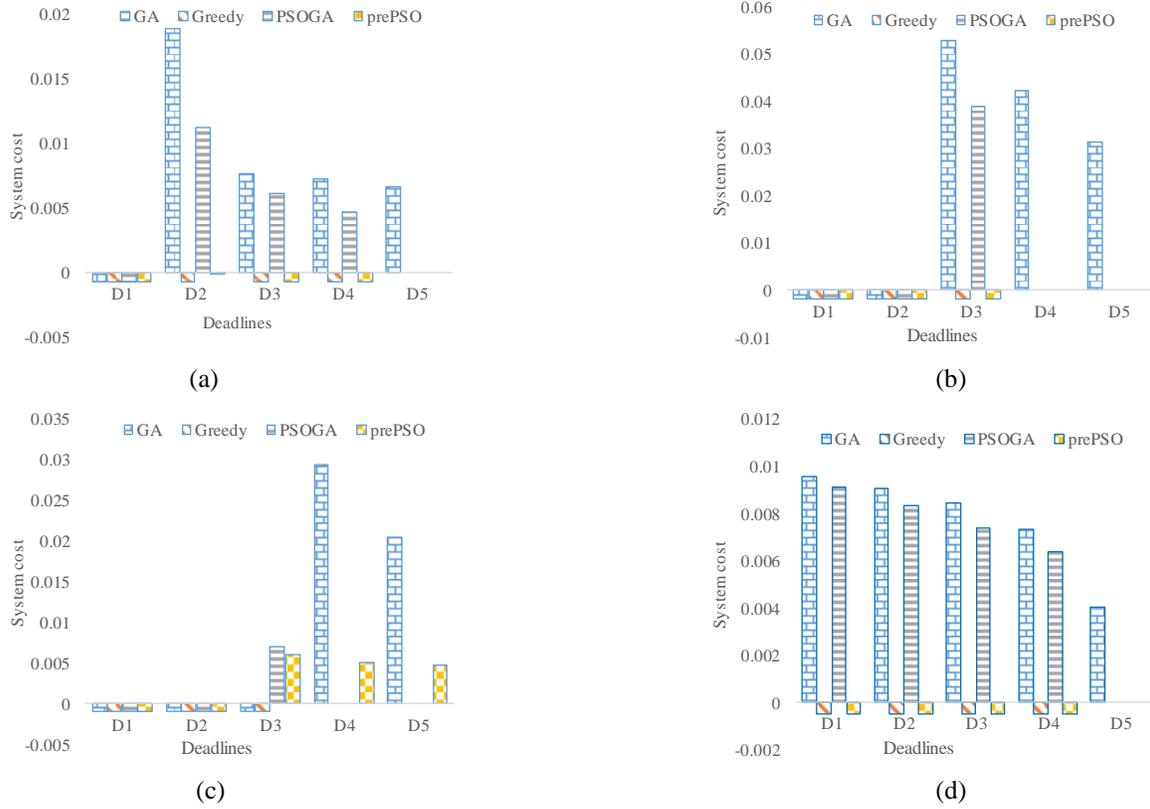

Fig. 8 The system cost of different strategies for three DNNs per end device: (a) AlexNet; (b) VGG19; (c) GoogleNet ; (d) ResNet101

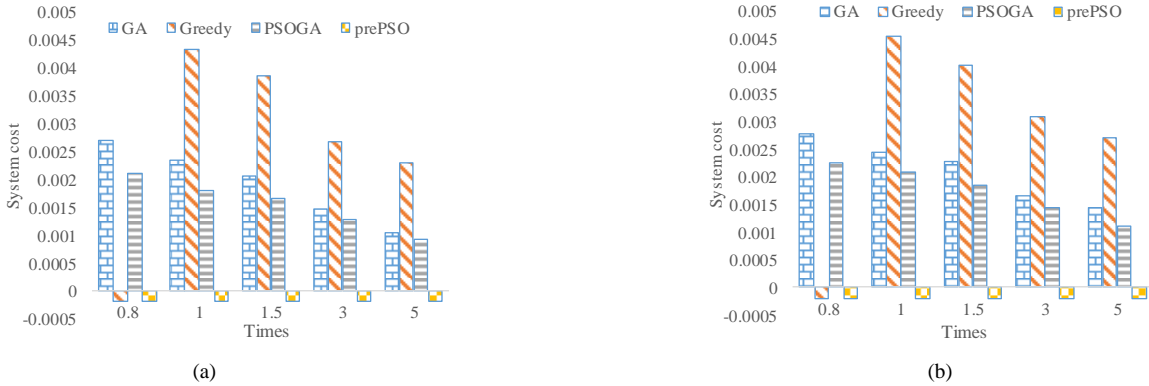

Fig. 9 The system cost of different strategies for one AlexNet per device in $D_2(G)$ with different updated computing power: (a) Edge; (b) Cloud

Fig. 8(a) shows The system cost of different strategies for three AlexNets per end device. As the total number of AlexNet layers increases, you can find that Greedy strategy becomes extremely unsuitable by comparing Fig. 8(a) with Fig. 7(a). Although PSO-GA has the best performance, it cannot offload all layers to the end devices to achieve 0 system cost until with $D_5(G)$ constraint. In Fig. 8(b), all offloading strategies except GA achieve 0 system cost after $D_3(G)$. It means that all layers in a AlexNet are offloaded to its original end device. In Fig. 8(d), PSO-GA and PSO both achieve the feasible solutions from $D_1(G)$ to $D_5(G)$. Greedy and prePSO can not find any feasible solution until in $D_5(G)$. prePSO compresses all the layers into one layer in a AlexNet, and offloads all layers in a AlexNet to its original end device when the corresponding deadline is loose enough.

In the follow-up experiments, we selected the representative AlexNet as the experimental subject. To observe the influence of the computing power of the edge server and cloud server on the performance of different strategies, we adjust the computing power of the edge server and cloud server based on the configurations for one AlexNet per device in $D_2(G)$. The updated computing power of the edge server and cloud server is {0.8, 1, 1.5, 3, 5} times that in the original settings.

Fig. 9 shows the system cost of different strategies for one AlexNet per device in $D_2(G)$ with different updated computing power of the edge server and cloud server. With the increase in the computing power of each edge server and cloud server, a layer on such server can be executed faster. This results in that more layers can be completed within the same time, and the layer computing cost will be reduced. Due to that more layers of AlexNet prefer to be offloaded to the edge servers, the increase in computing power of edge servers has more significant effect on the system cost than that in computing power of cloud servers. The system cost of different strategies with the increase in computing power of edge servers is 4% to 31% better than that with cloud servers.

Fig. 9(a) shows the system cost of different strategies for one AlexNet per device in $D_2(G)$ with different updated computing power of edge servers. AlexNet only have 11 layers, and the maximum amount of dataset transferring from a layer to its next layer is less that 1.1 MB in our benchmark test. From the experiments, we find that layer computing cost accounts for the majority of the total system cost for AlexNet. Most of layers are offloaded to the edge servers. For example, 69% of the number of layers and 51% of the amount of layer computing are on the edge servers. Fig. 9(b) shows the system cost of different strategies for one AlexNet per device in $D_2(G)$ with different updated computing power of cloud servers. It is obvious that the overall performance improvement of system cost is not as good as that in Fig. 9(a).

*D. Industrial Applications*

The package delivery application with unmanned aerial vehicles (UAV) swarms needs to rely on computer vision, whose core is DNNs. UAV swarms have limited battery capacity and are usually latency-intolerant. They should make real-time decisions with limited power to avoid collisions. With the hybrid computing environments, consisting of the cloud, edge and end devices, the larger layers (more amount of data) with high business intelligence are offloaded in the cloud, while the smaller layers (less amount of data) are offloaded at the edge and end devices. These three platforms collaborate with each other and execute the layers of DNNs with low system cost and latency. The offloading strategy based on PSO-GA proposed in this paper can effectively reduce the system cost within the specified deadline. Therefore, our strategy will extend the flight time of UAV swarms.

## VI. Conclusion

A cost-driven offloading strategy based on PSO-GA for DNN-based applications over the cloud, edge and end devices is proposed. The experimental results show that the offloading strategy effectively reduced the system cost within each DNN's corresponding deadline. With the looser deadline constraints, more layers can be offloaded to the cheaper servers under the same situation. This results in that the system cost decreases with the looser deadline. All layers can be offloaded in their original end devices with no data transmission, if their corresponding deadlines are loose enough. Moreover, the increase of server computing power has significant effect on the DNNs that their layer computing cost accounts for the majority of the total system cost.

In the future, the impact of bandwidth changes between two servers in different regions on the offloading strategies will be considered. In addition, each layer in the real environment has different price/performance ratios for different servers. Therefore, we will comprehensively optimize the system cost while considering the different price/performance ratios for different servers.